\documentclass[11pt,pdfwrite]{article}%
\usepackage{graphicx}
\usepackage{amsbsy}
\usepackage{amstext}
\usepackage{amsopn}
\usepackage{amsmath}
\usepackage{amsfonts}
\usepackage{amssymb}%
\begin{document}

\title{Certifying controls and systems software}
\author{Eric Feron\\School of Aerospace Engineering\\Georgia Institute of Technology\\\texttt{feron@gatech.edu}\\and \\Mardavij Roozbehani \\Laboratory for Information and Decision Systems\\Department of Aeronautics and Astronautics\\Massachusetts Institue of Technology\\mardavij@mit.edu }
\maketitle

\noindent\textbf{Abstract}

\noindent Software system certification presents itself with many challenges,
including the necessity to certify the system at the level of functional
requirements, code and binary levels, the need to chase down run-time errors,
and the need for proving timing properties of the eventual, compiled system.
This paper illustrates possible approaches for certifying code that arises
from control systems requirements as far as stability properties are
concerned. The relative simplicity of the certification process should
encourage the development of systematic procedures for certifying control
system codes for more complex environments.

\section{Introduction}

\begin{figure}
[ptb]
\begin{center}
\includegraphics[
height=6.4005in,
width=8.2814in
]%
{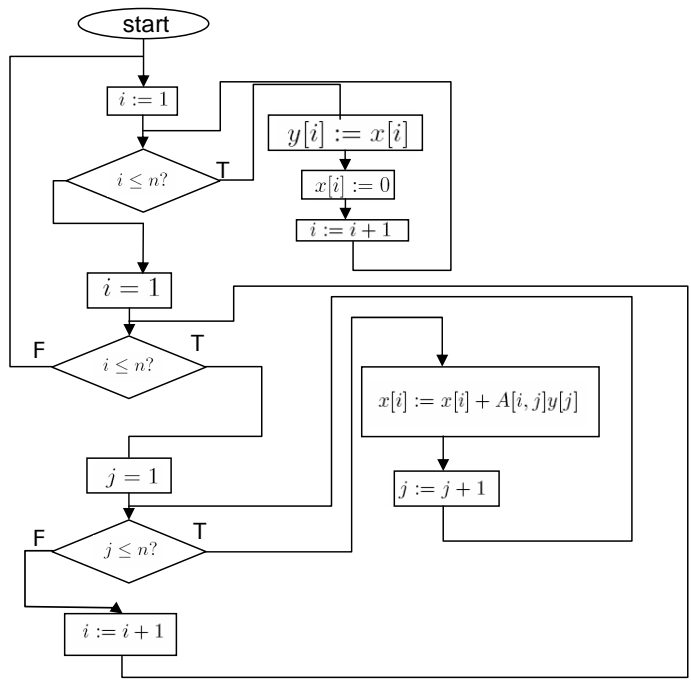}%
\caption{Program in flow-chart form}%
\label{program}%
\end{center}
\end{figure}

Discussions about analyzing software in a controls framework often oscillate
between trivial statements and feelings of doom dominated by undecidability
issues. Furthermore much of the intrinsic difficulty (or lack thereof) of
software analysis, even that designed by control systems engineers, hides
behind an intimidating language barrier, making tools and concepts developed
by computer scientists hard to reach by controls specialists, and vice-versa.
This paper concentrates on the following question: Given a properly designed
system (presumably a stable control system), what else needs to be proven to
convince the certification agency that the behavior is indeed appropriate? We
argue that the concept of \textquotedblleft proof of good
behavior\textquotedblright\ as currently taught in the control systems
curriculum mostly focuses at the level of \textquotedblleft system
specification\textquotedblright, meaning at a level where the system is
unambiguously defined, but does not constitute an executable code yet. By
means of example, consider the system
\begin{equation}
x_{k+1}=Ax_{k},\;\;k=0,1,\ldots\;\;\;x_{0}\in\mathrm{\mathbf{R}}^{n}%
,\;\;x_{0}^{T}x_{0}\leq1 \label{system}%
\end{equation}
and ask whether the state $x$ is always bounded. This question is usually
answered by checking various sufficient conditions such as (i) all eigenvalues
of $A$ have modulus strictly less than one, or (ii) there exists a symmetric,
positive definite matrix P satisfying the Lyapunov inequality
\[
A^{T}PA-P<0.
\]
While much of the control systems community would consider the job to be
\textquotedblleft done\textquotedblright, the fact is that the system (1) is
not considered \textquotedblleft executable\textquotedblright\ except for
high-level, non real-time environments such as MATLAB. The program described
in flowchart format in Fig. 1 is the accurate representation of what a
real-time, computer implementation of the system (1) might be. Discussing the
system (1) at the level of the flowchart shown in Fig. 1 might be dismissed by
many as an \textquotedblleft implementation issue\textquotedblright. This
opinion must, however, face the following facts: (i) Certification agencies
tend to look at \emph{all} levels of system implementation and not only the
specification level (1); (ii) the development of code is often not performed
by the same agent that developed the program specifications, thus introducing
doubt as to whether the specification has been faithfully implemented; and
(iii) providing the certification agency with \emph{code-level assurance} of
proper behavior does not require much effort, which is our main point today:
The next section of our paper discusses simple, generic philosophies when
establishing \textquotedblleft program proofs\textquotedblright, and argues in
favor of an \textquotedblleft instruction-by-instruction\textquotedblright%
\ analysis approach. This discussion is then applied to describe what
constitutes a \textquotedblleft proof\textquotedblright\ for the program shown
in Fig.~\ref{program}.%
We then conclude this article by suggesting easy extensions of our discussion.

\section{Proving computer programs: Gathering meaning first or not?}

When analyzing code, a central issue is the design of what is called the
collecting semantics. The collecting semantics describes how much information
from the original program must be retained and compiled in order to verify the
desired property (eg variable boundedness). The collecting semantics forms the
base model on which the analysis is conducted. Higher levels of semantic
collection allow one to define more compact models of the software execution,
but this task may also be more complex, since information must be collected
over several lines of code and then linked into a compact model. Thus, lower
semantic levels (like line-by-line analyses) are more desirable from the
standpoint of analyzer simplicity and adaptability. They may also improve
analysis readability, by linking the analysis closely to the code itself and
remembering that the value of any code analysis improves if it is more
readable. The process that favors high-level semantics collection first,
followed by the analysis of such semantics would correspond to taking the code
in Fig.~\ref{program},proving it matches the system
specification~(\ref{system}), and proving the specification~(\ref{system}) is
indeed stable using eigenvalue or other stability. Such an approach was
implemented, for example, in Cousot's ASTR\'EE analyzer~\cite{Feret:04,
blanchetcousotetal_pldi03, BlanchetCousotEtAL-Dagstuhl-3451-2003}, which was
used to support the certification process for several large commercial aircraft.

In contrast, the process that favors lower semantics levels, such as
``line-by- line'\ analyses would take the code in Fig.~\ref{program} and the
system specification~(\ref{system}) to build a proof that the code satisfies
the desired stability property.

The net result is a much more detailed analysis of the code at a level that
stands much closer to its eventual implementation. Another key observation is
to realize that once written, this proof does not require the system
specification~\ref{system} to be understood and independently verified. A key
tool for line-by-line analysis may be found in~\cite[Ch.7]{Pel:01}, where the
author describes a technique for line-by-line analysis by means of invariants:
In this analysis, which dates back to the 1960's, each line of code
corresponds to either a test or an assignment. A test has one entry channel
and two exit channels (one for the case when the test is true, and one for
when the test is false). An assignment has one input channel and one output
channel, and consists of a variable update by means of a domainspecific
operation. In the code described in Fig.~\ref{program}, tests are shown in
diamond-shaped boxes, while assignments are shown in rectangular boxes. For
each line of code, two invariant properties of the code state are guessed: The
pre-instruction invariant describes a set to which the code state always
belongs prior to instruction execution. The post-instruction invariant(s)
describe set(s) to which the variables always belong after the execution of
the instruction. Given two consecutive instructions named \textbf{1} and
\textbf{2}, if the post-instruction invariant of \textbf{1} is the same as the
pre-instruction invariant of \textbf{2}, then the two instructions can be
composed with the guarantee that the pre-instruction of instruction \textbf{1}
implies the post-instruction invariant of instruction \textbf{2}. If such
compatibility conditions hold over the whole program, then it is possible to
use such mechanisms to prove that an entire program is ``correct''. The appeal
of such a method is that the elements of proof are distributed throughout the
code, and may be independently verified instruction by instruction either
manually or using the help of a computer, with no need to understand the whole
``meaning'\ of the program. 2 Proving the implementation of a linear system
Consider now the code described in Fig.~\ref{program}. We use our previous
discussion to show that the correctness of such a code may be established on a
line-by-line basis using ellipsoidal invariants. In the context of this
section, we will say the program is correct if, for bounded initial
conditions, the variables inside the program are all bounded. This observation
becomes trivial once we recognize that (i) the proper state-space for the code
consists of $x$, $y$ and the index variables $i$ and $j$ and (ii) all
operations involving the computer variables $x$ and $y$ are linear in $x$ and
$y$. From a theoretical perspective, describing the code behavior by means of
ellipsoidal invariants adds little more to the story than simply describing it
by means of (quadratic) Lyapunov functions. However, we find that ellipsoids
are easier to manipulate, in that they eliminate inconvenient singular matrix
inverse computations.

\section{Behavior of state variables}

The solution to the problem begins with realizing that the relevant
state-space includes both $x$ and $y$, and that the transformations on $x$ and
$y$ are linear in $x$ and $y$. For example, the instruction
\[
y[i] :=x[i]
\]
is a linear transformation that may also be written
\begin{equation}
\left[
\begin{array}
[c]{c}%
y\\
x
\end{array}
\right]  = \left[
\begin{array}
[c]{cc}%
I-e_{ii} & e_{ii}\\
0 & I
\end{array}
\right]  \left[
\begin{array}
[c]{c}%
y\\
x
\end{array}
\right]  = S_{i}\left[
\begin{array}
[c]{c}%
y\\
x
\end{array}
\right]  \label{transf1}%
\end{equation}
where $e_{ij}$ is a matrix made of zeros everywhere, except at the entry on
the $i$th row and $j$th column which is 1. Likewise, the instruction
\[
x[i] := 0
\]
is the linear transformation
\begin{equation}
\left[
\begin{array}
[c]{c}%
y\\
x
\end{array}
\right]  = \left[
\begin{array}
[c]{cc}%
I & 0\\
0 & I-e_{ii}%
\end{array}
\right]  \left[
\begin{array}
[c]{c}%
y\\
x
\end{array}
\right]  = P_{i}\left[
\begin{array}
[c]{c}%
y\\
x
\end{array}
\right]  \label{transf2}%
\end{equation}
and the instruction
\[
x[i] := x[i] + A[i,j]y[j]
\]
is the linear transformation
\begin{equation}
\left[
\begin{array}
[c]{c}%
y\\
x
\end{array}
\right]  = \left[
\begin{array}
[c]{cc}%
I & 0\\
A[i,j]e_{ij} & I
\end{array}
\right]  \left[
\begin{array}
[c]{c}%
y\\
x
\end{array}
\right]  = T_{ij}\left[
\begin{array}
[c]{c}%
y\\
x
\end{array}
\right]  \label{transf3}%
\end{equation}
After a full cycle, the net effect is for $x$ and $y$ to have gone through the
multiplexed operation:
\begin{equation}
\left[
\begin{array}
[c]{c}%
y\\
x
\end{array}
\right]  = \left[
\begin{array}
[c]{cc}%
0 & I\\
0 & A
\end{array}
\right]  \left[
\begin{array}
[c]{c}%
y\\
x
\end{array}
\right]  = A_{1}\left[
\begin{array}
[c]{c}%
y\\
x
\end{array}
\right]  \label{net-loop}%
\end{equation}

\section{Ellipsoids}

We use ellipsoidal invariants before and after each instruction in the
program. Ellipsoids (centered around the origin) are usually defined as
\begin{equation}
\left\{  z \in\mathrm{\mathbf{R}}^{n} \; \left|  \; z^{T}Pz \leq1\right.
\right\}  , \label{ellipses}%
\end{equation}
where $P$ is a symmetric, positive semidefinite matrix. This definition
captures all ellipsoids whose volume is strictly positive, including infinite
when $P$ is singular. Yet it fails to capture all the finite ellipsoids of
interest in this discussion, including degenerate ellipsoids (eg flat,
``pancake-like'\ ellipsoids). For this reason, we prefer to define the
ellipsoid ${\mathcal{E}}_{R}$ as
\begin{equation}
{\mathcal{E}}_{R} = \left\{  z \in\mathrm{\mathbf{R}}^{n}\;\; \left[
\begin{array}
[c]{cc}%
R & z\\
z^{T} & 1
\end{array}
\right]  \geq0\right\}  , \label{ellipses2}%
\end{equation}
where $R$ is a symmetric, positive semidefinite matrix. It is easy to see, by
means of Schur complements~\cite{BEFB:94}, that~(\ref{ellipses2})
and~(\ref{ellipses}) refer to the same ellipsoid if $P$ is invertible and
$P^{-1} = R$. On the other hand, a singular $R$ indicates a bounded,
degenerate ellipsoid.

\section{Behavior of ellipsoids under linear tranformations}

The behavior of ellipsoids under linear transformation is well-known (see, for
example~\cite{Kur:96} for example). Consider the map
\begin{equation}
z = Tv \label{map}%
\end{equation}
where $T$ is a matrix. Then the bounded ellipsoid ${\mathcal{E}}_{R}$ becomes
the bounded ellipsoid ${\mathcal{E}}_{TRT^{T}}$ through the
transformation~(\ref{map}). Thus, if~(\ref{map}) corresponds to a program
instruction and ${\mathcal{E}}_{R}$ is a pre-instruction assertion, then
${\mathcal{E}}_{TRT^{T}}$ is a valid post-instruction invariant.

\section{Instruction-level annotation of control programs with ellipsoidal
invariants}

With this simple design rule, we can now properly annotate the program with
ellipsoidal invariants at the instruction level. The dimension of the
ellipsoids is $2n$ since the state include both $y$ and $x$.%

\begin{figure}
[ptb]
\begin{center}
\includegraphics[
height=4.881in,
width=5.3973in
]%
{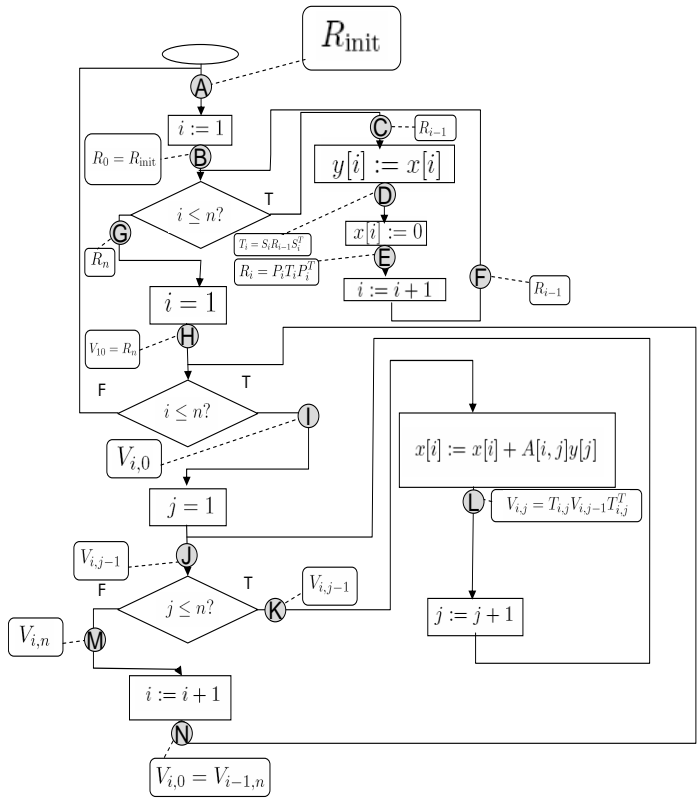}%
\caption{Program annotated with invariants}%
\label{annotated_program}%
\end{center}
\end{figure}
The annotation is shown in Fig.~\ref{annotated_program}, and relies on the
definitions of $S_{i}$, $P_{i}$ and $T_{ij}$ given in~(\ref{transf1}%
),~(\ref{transf2}),and~(\ref{transf3}). For the sake of simplicity, all
invariants are summarized by the corresponding symmetric, semidefinite
matrices. Based on our prior discussion, the coherence of these invariants is
trivial throughout the program, except (i) during the initialization step
(computation of Rinit) and (ii) during the transition from the link denoted
$N$ to the link denoted $A$, that is, at the end of the outermost program
loop: The ellipsoid ${\mathcal{E}}_{V_{nn}}$ must be contained in the
ellipsoid ${\mathcal{E}}_{R_{\mathrm{init}}}$ . For these two properties to
hold we pick $R_{\mathrm{init}}$ according to the following observations: As
noted earlier, the loop that begins an d ends at the location A performs the
following overall matrix iteration~(\ref{net-loop}) on $y$ and $x$. Since $A$
is stable, so is $A_{1}$ and
\begin{equation}
V_{nn}=A_{1}R_{\mathrm{init}}A_{1}^{T}. \label{net-ellips}%
\end{equation}
Thus we can always find $P>0$ such that
\begin{equation}
A_{1}^{T}PA_{1}-P<0. \label{net-lyap}%
\end{equation}
Let us define
\[
R_{\mathrm{init}}=\alpha P^{-1}%
\]
where $\alpha$ is a positive scaling parameter. By virtue of
Eqs.~(\ref{net-lyap}) and~(\ref{net-ellips}),
\[
{\mathcal{E}}V_{nn}\subseteq{\mathcal{E}}_{R_{\mathrm{init}}}%
\]
whatever the value of $\alpha$. We then pick $\alpha$ to scale ${\mathcal{E}%
}_{R_{\mathrm{init}}}$ so as to contain all allowed initial values of $x$ and
$y$. Assume for example $|x_{i}|\leq1$, $y_{i}=0$, $i=1,\leq,n$ at program
initialization. Then $x^{T}x+y^{T}y\leq n$, that is $x$ and $y$ are contained
in the ball ${\mathcal{E}}_{nI}$. Then let $\sigma_{\mathrm{max}}$ be the
largest eigenvalue of $P$ satisfying~(\ref{net-lyap}). $\sigma_{\mathrm{max}}$
is necessarily positive, and the unit ball ${\mathcal{E}}_{I}$ is included in
the ellipsoid ${\mathcal{E}}_{\sigma_{\mathrm{max}}P^{-1}}$ . Let
$\alpha=n\sigma_{\mathrm{max}}$. Then the ball ${\mathcal{E}}_{nI}$ is
contained in the ellipsoid ${\mathcal{E}}_{\alpha P^{-1}}={\mathcal{E}%
}_{R_{\mathrm{init}}}$. The annotation process is now complete. To obtain the
range over which all variables live in, simply compute the union of all
computed ellipsoids above, or compute, say, the minimum volume (or trace)
ellipsoid containing all ellipsoids above, or even siompler the minimum-sized
ball containing all these ellipsoids.

\newpage

The following code, written in MATLAB automates this annotation process for a
given $A$ matrix
\begin{verbatim}
%This code does the documentation automatically...
% for a 2x2 system
%Pick an example
A = [0 1;-0.1 -0.2]
Q=eye(4);
A1 = [zeros(2,2) eye(2);zeros(2,2) A]
%P computation
P = dlyap(A1',Q);
%scaling of P
u = max(eig(P));
n = 2;
alpha = 1/u/n/2;
P = alpha*P
%annotation beginning
Rinit = P^(-1);
R0 = Rinit
% i=1
S1 = [eye(2,2) zeros(2,2);zeros(2,2) eye(2,2)];
S1(1,1)=0;S1(1,3)=1;
T1 = S1*R0*S1'
P1 = [eye(2,2) zeros(2,2);zeros(2,2) eye(2,2)];
P1(3,3) =0;
R1 = P1*T1*P1'
% i=2
S2 = [eye(2,2) zeros(2,2);zeros(2,2) eye(2,2)];
S2(2,2)=0;S2(2,4)=1;
T2 = S2*R1*S2'
P2 = [eye(2,2) zeros(2,2);zeros(2,2) eye(2,2)];
P2(4,4) =0;
R2 = P2*T2*P2'
V10 = R2
% i=1 j=1
T11 = [eye(2,2) zeros(2,2);zeros(2,2) eye(2,2)];
T11(3,1)= A(1,1);
V11 = T11*V10*T11'
% i=1 j=2
T12 = [eye(2,2) zeros(2,2);zeros(2,2) eye(2,2)];
T12(3,2)= A(1,2);
V12 = T12*V11*T12'
% i=2 j=1
T21 = [eye(2,2) zeros(2,2);zeros(2,2) eye(2,2)];
T21(4,1)= A(2,1);
V21 = T21*V12*T21'
% i=2 j=2
T22 = [eye(2,2) zeros(2,2);zeros(2,2) eye(2,2)];
T22(4,2)= A(2,2);
V22 = T22*V21*T22'
%end of loop
% This line checks that indeed, the ellipsoid defined by V22
% is contained in that defined by Rinit;
% The computed eigenvalues ought to be all negative
display('Eigenvalues')
[vV22,eV22]=eig(V22-Rinit)
%quit
\end{verbatim}

\section*{Conclusion}

This paper has shown one process by which it is possible to carry control
system specification certificates down to their software implementation. This
process has been shown to be relatively straightforward. We hope to have
convinced the reader that this process may be applied to more sophisticated,
computer-controlled systems, and that it will eventually lead to the
development of more rigorous, yet flexible, embedded software development
practices, from its specification down to its operation.

\section*{Acknowledgements}

Many thanks to Pete Manolios, Alexandre Megretski, Olin Shivers, Arnaud Venet
and Daron Vroon for useful discussions. Many thanks to Fernando Alegre for
editing the MATLAB script.

"This research was supported by the National Science Foundation under grant
CSR/EHS 0615025, the Dutton/Ducoffe professorship at Georgia Tech and the
Massachusetts Institute of Technology."

\bibliographystyle{alpha}
\bibliography{bibtex_database}

\end{document}